\documentclass[12pt,preprint]{aastex} 

\shorttitle{The dusty disk around VV~Ser}
\shortauthors{Alonso-Albi et al.}

\begin{document}
   \title{The dusty disk around VV~Ser}

\author{T. Alonso-Albi\altaffilmark{1}, A. Fuente\altaffilmark{1}, 
R. Bachiller\altaffilmark{1}, R. Neri\altaffilmark{2}, P. Planesas\altaffilmark{1}, L. Testi\altaffilmark{3, 4}}

\altaffiltext{1}{Observatorio Astron\'omico Nacional, Aptdo. Correos 112, 28803 Alcal\'a de Henares (Madrid), Spain}
\altaffiltext{2}{Institut de Radio Astronomie Milimetrique, 300 Rue de la Piscine, Domaine Universitaire de Grenoble, F-38406 St. Martin d'H\`{e}res, France}
\altaffiltext{3}{INAF-Osservatorio Astrofisico de Arcetri, Largo Enrico Fermi 5, I-50125 Firenze, Italy}
\altaffiltext{4}{ESO, Karl Schwarzschild str.2, D-85748 Garching bei Muenchen, Germany}

\begin{abstract}

We have carried out observations at millimeter and centimeter wavelengths towards VV~Ser using the Plateau de Bure Interferometer and the Very Large Array. This allows us to compute the SED from near infrared to centimeter wavelengths. The modeling of the full SED has provided insight into the dust properties and a more accurate value of the disk mass.

The mass of dust in the disk around VV~Ser is found to be about $4$~$10^{-5}$ M$_\odot$, i.e. 400 times larger than previous estimates. Moreoever, the SED can only be accounted for assuming dust stratification in the vertical direction across the disk. The existence of small grains (0.25--1 $\mu$m) in the disk surface is required to explain the emission at near- and mid-infrared wavelengths. The fluxes measured at millimeter wavelengths imply that the dust grains in the midplane have grown up to very large sizes, at least to some centimeters.

\end{abstract}


\section{Introduction}

Herbig Ae and Be (HAEBE) stars are intermediate mass (M$\sim$2-8 M$_\odot$) pre-main sequence objects. Since these objects share many characteristics with high-mass stars (clustering, PDRs), and some of them are much closer, the detection and further study of the circumstellar disks around these stars is crucial for the understanding of the massive star formation process. In addition, Herbig Ae (HAe) stars are the precursors of Vega-type systems and therefore the comprehension of the disk evolution around HAe stars is a key stone in planet formation studies.

Theoretical and observational efforts have been made in recent years towards the understanding of the disk occurrence and evolution in HAEBE stars \citep{mil01, mee01, ack05, fue03, fue06}. The observed disks reveal a large source to source variation that suggests an evolution from young flaring gas-rich disks to tenuous debris disks (see, e.g., \cite{gra05, ram06, gee07b, mar07, eis07, kal07}). Theoretical models predict grain growth and dust settling in the disk midplane that causes the optical depth of the disk to drop and allows the stellar UV radiation to penetrate deep into the circumstellar disk to photo-evaporate the external layers \citep{dul04}. However the observational evidence for this process is limited. Our comprehension of the dust properties in the disks is mainly based on the dust continuum emission and spectral energy distribution (SED) modeling. Dust thermal emission is usually optically thick at infrared (IR) wavelengths and only gives information on the temperature and properties of the small grains in the disk surface. Using the spectral index of the dust emission in the mm range is the only way we can probe the properties of the dust in the midplane \citep{ale01}.

In this Paper we present mm and cm wavelength observations and the modeling of the disk around VV~Ser. VV~Ser is an Herbig A0e star, 260 pc distant, surrounded by a nearly edge-on disk previously detected in near and mid IR images \citep{pon07}. As a UX Orionis variable star, it has frequent and irregular extinction events that could be interpreted as produced by clumps in the inner rim close to the star \citep{nat01}. \cite{pon07} modeled the near and far IR SED and showed that it is consistent with a self-shadowed disk with a radius of 50 AU and a mass of 8~10$^{-8}$ M$_\odot$. We have modeled the whole SED of VV~Ser adding our new mm and cm observations. We find that all available photometry can be explained as arising from a dusty disk 400 times more massive than previously determined, in which grain growth has proceeded to sizes of about 1~cm. 

\section{Observations}

Continuum observations at 115.3 GHz and 230.5 GHz and spectroscopic observations of the $^{12}$CO 1$\rightarrow$0 and $2\rightarrow1$ lines were carried out with the IRAM - Plateau de Bure Interferometer\footnote{IRAM is supported by INSU/CNRS (France), MPG (Germany), and IGN (Spain).} (PdBI) in April, 2006. The spectral correlator setup was adjusted to observe CO 1$\rightarrow$0
and CO 2$\rightarrow$1 lines, providing a spectral resolution
of ~40 kHz over a bandwidth of 20 MHz. Additional observations in continuum at 3~mm were obtained in August, 2007. 
The synthesized beams are 4.0\arcsec x2.8\arcsec  at 115.3 GHz, and 1.7\arcsec x0.8\arcsec  at 230.5 GHz. The absolute flux density was determined from measurements on 3C273, 1741-038, MWC349, and 3C84. Continuum observations at cm wavelengths were obtained in November, 2005, using the NRAO Very Large Array (VLA) in its D configuration. The synthesized beams were 2.2\arcsec x1.7\arcsec , 5.2\arcsec x3.2\arcsec , and 17.8\arcsec x8.0\arcsec  at 44 GHz, 22 GHz, and 8 GHz respectively. 
The calibration sources were 0137+331 and 18518+00355 in the 44 GHz observation, 18518+00355 at 22 GHz, and 1851+005 at 8 GHz. 
All the images are centered at the star position RA: 18h 28m 47.86s, DEC: +00$^\circ$ 08' 40.0\arcsec  (J2000).

\clearpage
\begin{figure}
\centering
 \includegraphics[width=.45\textwidth, angle=-90, viewport=0 100 500 600]{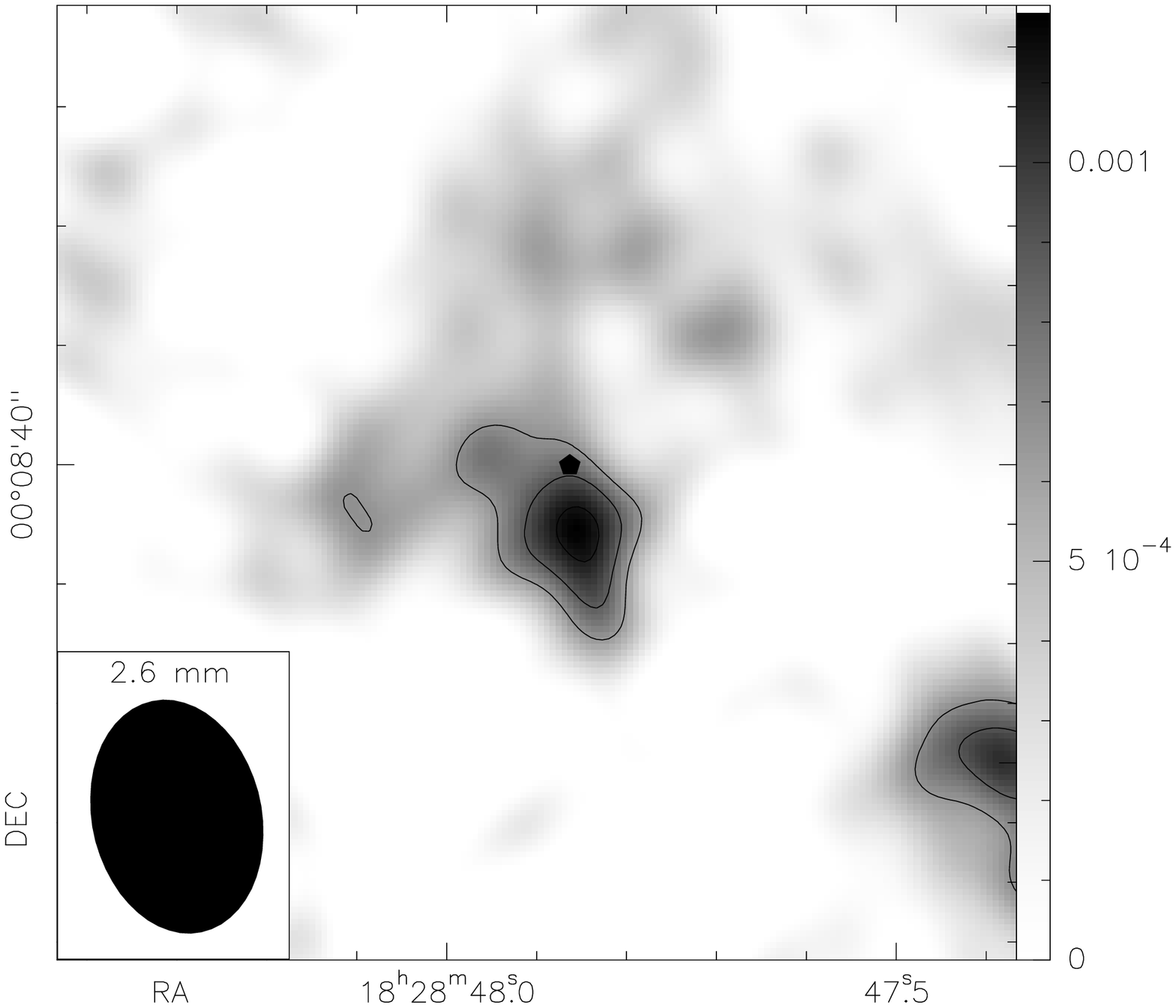}\hspace*{7mm}
 \includegraphics[width=.45\textwidth, angle=-90, viewport=0 100 500 600]{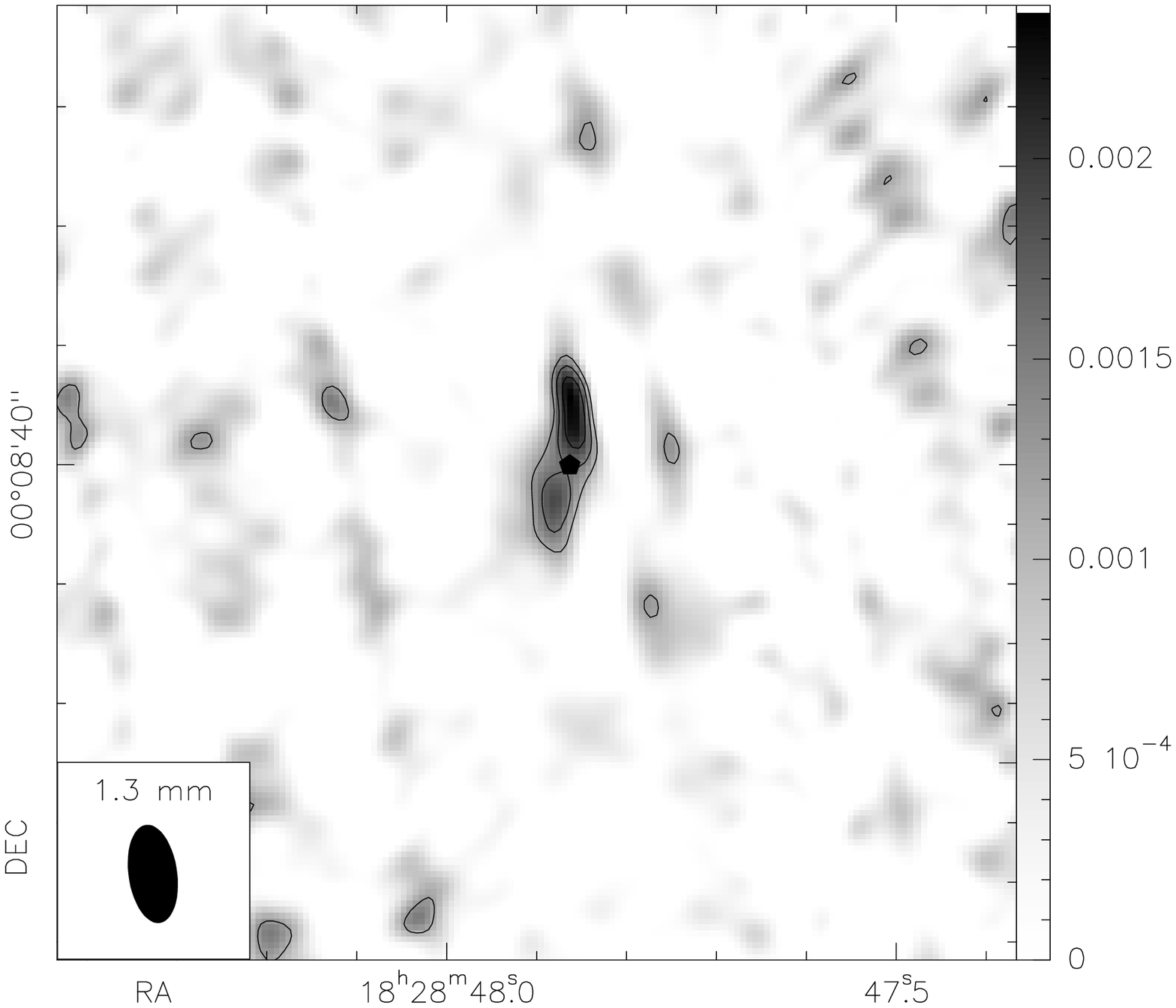}
 \smallskip
\caption{
Left: Map at 2.6~mm obtained with the PdBI. Countours are  in steps of 0.22~mJy/beam (1$\sigma$), from 0.66 to 1.1~mJy/beam. Right: Map at 1~mm obtained with PdBI. Contours are in steps of 0.4~mJy/beam (1$\sigma$), from 1.2 to 2.0~mJy/beam.}
\label{fig:0}
\end{figure}
\clearpage

\section{Results}
A summary of our mm and cm observations is presented in Table~\ref{tab:1}. We tentatively detected VV~Ser at 7~mm using VLA with a flux of 0.35~mJy. We did not detect continuum emission at 3.6 and 1.3~cm with a sensitivity (3$\sigma$) of 0.15 and 0.25~mJy/beam, respectively. In our model we adopted the upper limit of 0.09~mJy/beam at 3.6~cm from \cite{ski93}, since it is more restrictive.

We detected (5$\sigma$) the disk around VV~Ser at 2.6~mm and 1.3~mm using the PdBI. Our 2.6mm image shows evidence of the existence of an extended emission component that has been filtered out by the interferometric observations (see Fig.~1).
This is consistent with the large nebula detected by \cite{pon07} in the Spitzer image at 70~$\mu$m. 
The detected peak emission (over 4$\sigma$) at 2.6~mm and 1.3mm is consistent with the star position, since the offset between the star position and the peak emission (1\arcsec at 3~mm and 0.8\arcsec at 1~mm) is less than half the size of the synthesized beam. Furthermore, the sizes of the emitting region in the two images are smaller than the synthesized beams. Consequently we consider that the detected emission is mainly associated with the star+disk system. Possibly a fraction of the emission at 2.6 and 1.3~mm is due to free-free emission arising from the stellar wind. To substract this contribution we have extrapolated the upper limit to the emission at 3.6~cm to mm wavelengths, assuming a theoretical spectral index of 0.6. Even in this pessimistic case we found an emission excess at 1.3 and 2.6~mm ($\sim$50$\%$) that is very likely due to the dusty disk (see Table~\ref{tab:1}).

We did not detect the disk emission in the $^{12}$CO 1$\rightarrow$0 and 2$\rightarrow$1 images. Our $^{12}$CO 1$\rightarrow$0 map has a sensitivity of 14~mJy (0.3~K) in a channel-width of 1~km/s. As will be discussed later this limit is consistent with a small optically thick CO disk.



\clearpage
\begin{table}
\caption{Summary of VV~Ser observations.}
\begin{tabular}{llll}
\hline\noalign{\smallskip}
Wavelength (cm) & Flux (mJy) & Beam (\arcsec ) & Disk flux (mJy)\tablenotemark{(a)} \\
\noalign{\smallskip}\hline\noalign{\smallskip}
3.552 & $<0.09$ \tablenotemark{(b)} & 0.3 & \\
3.544 & $<0.15$ & 17.8 x 8.0 & \\
1.335 & $<0.25$ & 5.2 x 3.2 & \\
0.692 & 0.35 $\pm$ 0.1 \tablenotemark{(c)} & 2.2 x 1.7 & 0.11 \\
0.260 & 1.10 $\pm$ 0.22 & 4.0 x 2.8 & 0.67 \\
0.130 & 2.1 $\pm$ 0.4 & 1.7 x 0.8 & 1.44 \\


\noalign{\smallskip}\hline
\end{tabular}
\label{tab:1}       

\tablenotetext{(a)}{Estimated from model by substracting free-free contribution.}
\tablenotetext{(b)}{From \cite{ski93}. Flux limits are 3$\sigma$ values.}
\tablenotetext{(c)}{Flux errors are 1$\sigma$ values.}
\end{table}
\clearpage

\section{Modeling the SED}

To have a deeper insight into the structure of the disk and dust properties, we have fitted the whole SED using the passive irradiated circumstellar disk model developed by \cite{dul01}. 
Fluxes from 0.2 to 70~$\mu$m have been taken from \cite{pon07}, and our cm and mm data have been used to complete the SED at longer wavelengths.
The model is based on the original Paper by \cite{chi97}, and describes the dust around the circumstellar disk of an HAe star by means of three layers (one interior in the midplane and two surface layers) in radiative equilibrium. In addition, an inner rim located at the dust evaporation radius is added by \cite{dul01} to account for the near IR bump observed in the SEDs of some HAe stars.

We have developed code to account for different grain populations characterized by the silicate/graphite mixture, the maximum grain size, and the grain size distribution slope. The optical properties of the astronomical silicate (olivine and orthopyroxene) and graphite are assumed to be those from \cite{dra84} and \cite{lao93}. We use the Mie scattering code from Bohren and Huffman (1983) to calculate the absorption cross sections considering that the grains are homogeneous isotropic spheres. For graphite the 1/3-2/3 approximation is used \citep{dra93}. The size distribution is taken to be a power function law $n(a)~=~n_0~a^{-p}$, where $a$ is the grain radius, $n_0$, a normalization factor that depends on the assumed maximum grain size in the distribution, and $p$, a parameter in the range 2.5-3.5. The dust opacity is extrapolated for each type of grain for wavelengths beyond 1000~$\mu$m, where there is no available information for the refractive index. The assumed density for silicate is 3.50~g~cm$^{-3}$, and 2.24~g~cm$^{-3}$ for graphite. The corresponding sublimation temperatures are 1500~K and 2000~K. The dust in the rim is considered to have the same properties as the dust in the surface layer, but different from the dust in the midplane.

The spectral type of this source ranges from B6 \citep{her04} to A2 \citep{cha88}, that corresponds to effective temperatures between 8800 and 13500 K. Spectral type A0 determined by high resolution spectra \citep{mor01} is probably the best estimate considering the irregular light variations exhibited by the star. We have adopted therefore this spectral type and an effective temperature of 10000 K (see Table 2). The distance of 260 pc corresponds to the Serpens molecular cloud \citep{str96}. The derived luminosity is 49 $\pm$ 5 L$_\odot$ \citep{pon07}.

\begin{figure*}
\includegraphics[viewport=20 60 480 700, width=0.9\textwidth, height=1.0\textwidth, angle=-90]{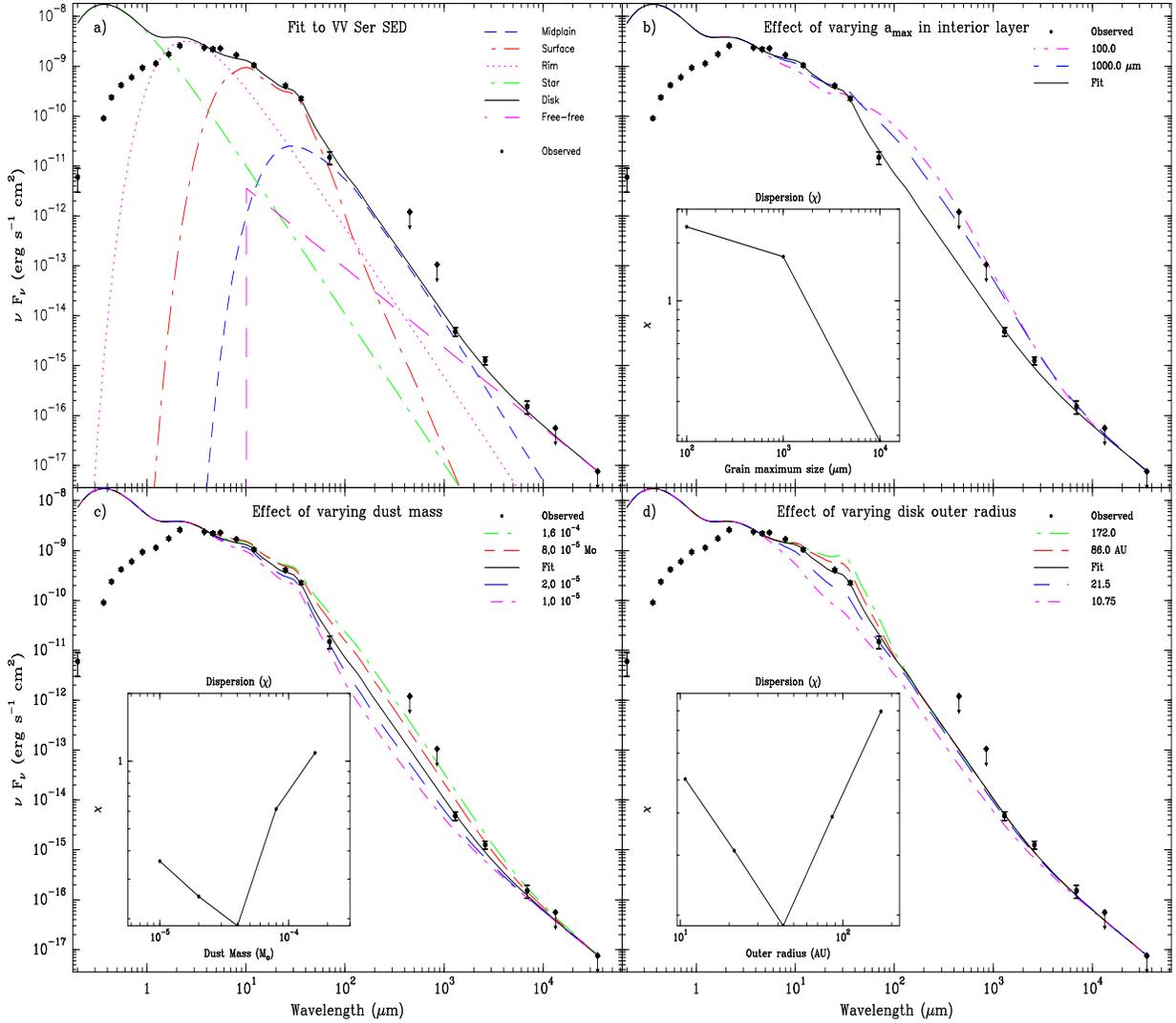}
\caption{{\bf (a)} The fitted SED is shown with all the model contributions: interior layer (blue), surface layer (red), inner rim (pink), star (green), free-free (pink), total sum of previous contributions (gray). Observed points at IR wavelengths ($<$1~mm) are taken from the literature \citep{pon07}. The points at mm and cm wavelenghts correspond to this work. 
{\bf (b)} Model predicted SED using all the parameters shown in Table 2 except the value of the maximum grain size in the interior layer (a$_{max}$) that is let to vary between 100~$\mu$m, 1~mm and 1~cm. The inset shows the obtained dispersion ($\chi$) as a function of a$_{max}$. We have not made calculations for a$_{max}>$1~cm. {\bf (c)} The same as (b) but varying the disk mass. The results for disk masses of 1.0E-5, 2.0E-5, 4E-5 (fitted value), 8E-5 and 1.6E-4 M$_\odot$ are shown. (d) The same as (b) but varying the outer radius. The results for outer radius of 10.75, 21.5, 43 (fitted value), 86 and 172 AU are shown.}
\label{fig:1b}
\end{figure*}

In our model we fix the stellar parameters and the rim temperature (and consequently, the rim radius), whereas the disk parameters are allowed to vary independently. To evaluate the fit we use a dispersion ($\chi$) parameter defined as $\chi~=~\sqrt{\sum{((F_{observed}-F_{modeled})/F_{observed})^2}/n}$, where F is the flux and n the number of detections in the SED beyond 3~$\mu$m, without considering the upper limits. Since the flux is higher at near- and mid- IR wavelengths, this dimensionless parameter is the most suitable because it gives the same weight to all the fluxes in the fit (specially our mm and cm observations), regardless of the different absolute flux values. We use a dispersion minimization algorithm to find the best solution. The resulting parameters are listed in Table~\ref{tab:1b} and the predicted SED is shown in Fig.~\ref{fig:1b}.

\clearpage

\begin{figure}
\centering
\includegraphics[width=.45\textwidth, angle=-90, viewport=0 100 500 400]{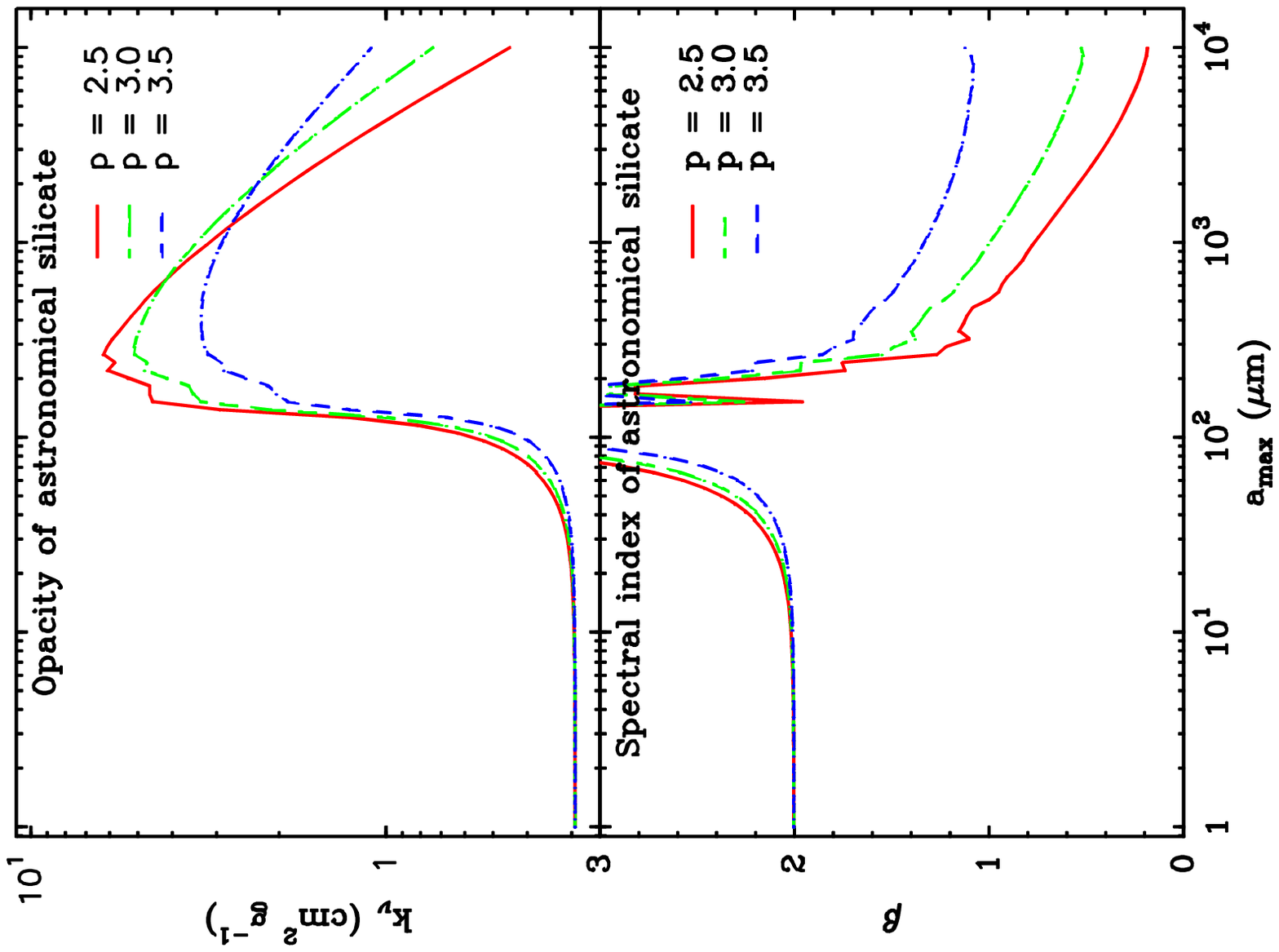}\hspace*{7mm}
\includegraphics[width=.45\textwidth, angle=-90, viewport=0 100 500 400]{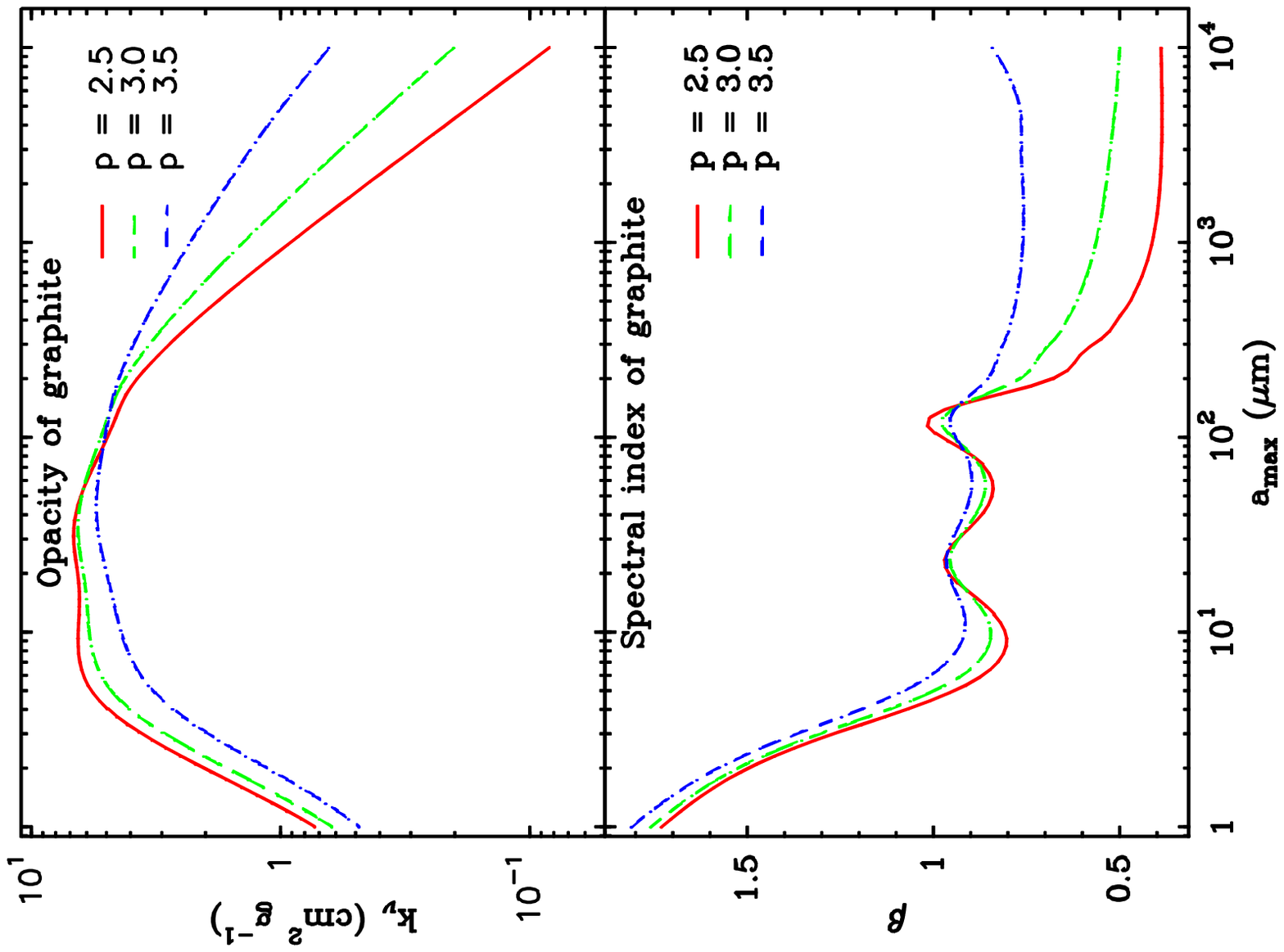}
\caption{Opacity and spectral index of astronomical silicate and graphite at 1 mm as function of the maximum grain size, for different values of the grain size distribution slope. Spectral index is calculated between 0.8 and 1 mm.}
\label{fig:2a}
\end{figure}

\begin{figure*}
\includegraphics[width=0.5\textwidth, height=0.36\textwidth]{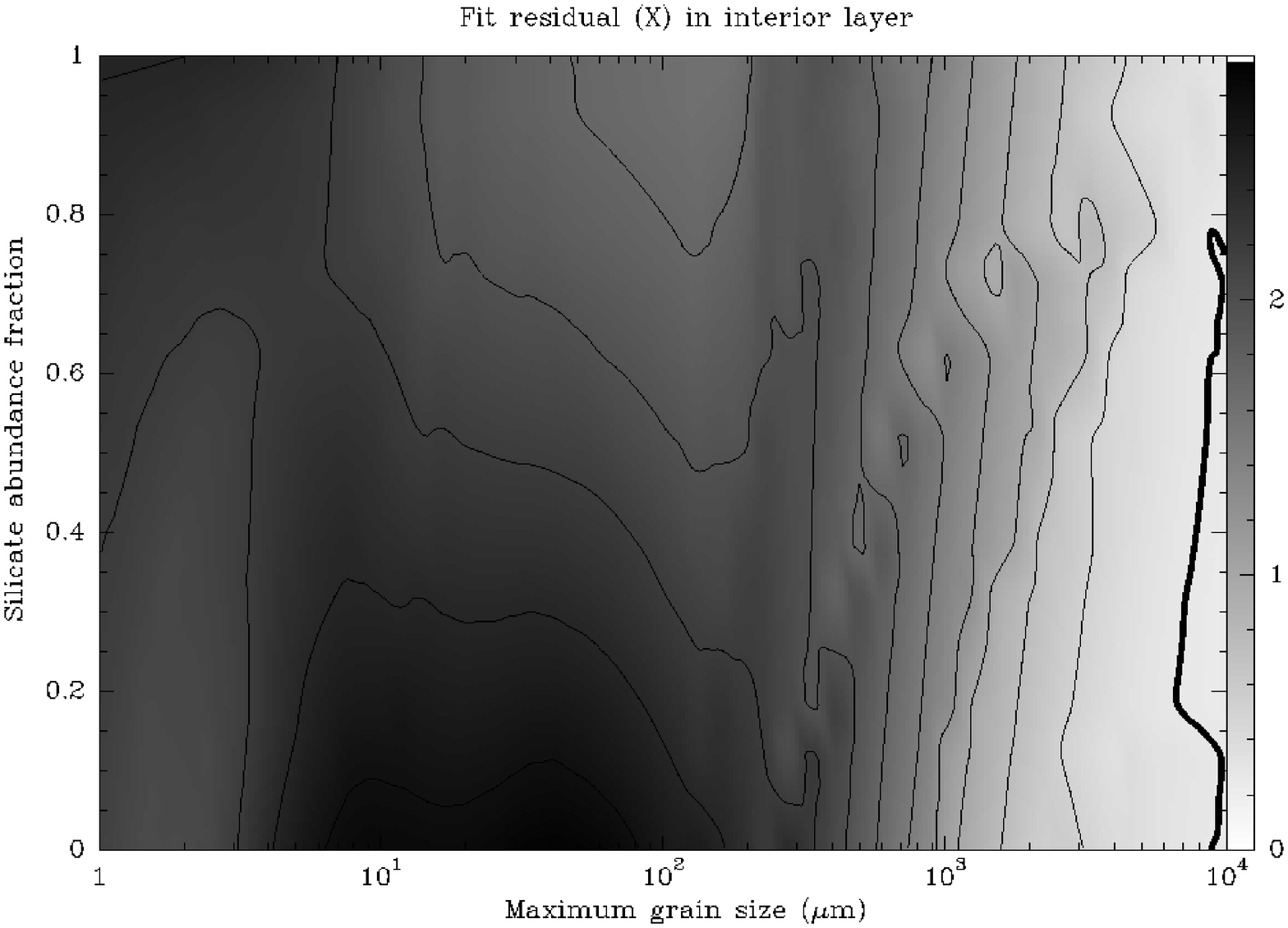}
\includegraphics[width=0.5\textwidth, height=0.36\textwidth]{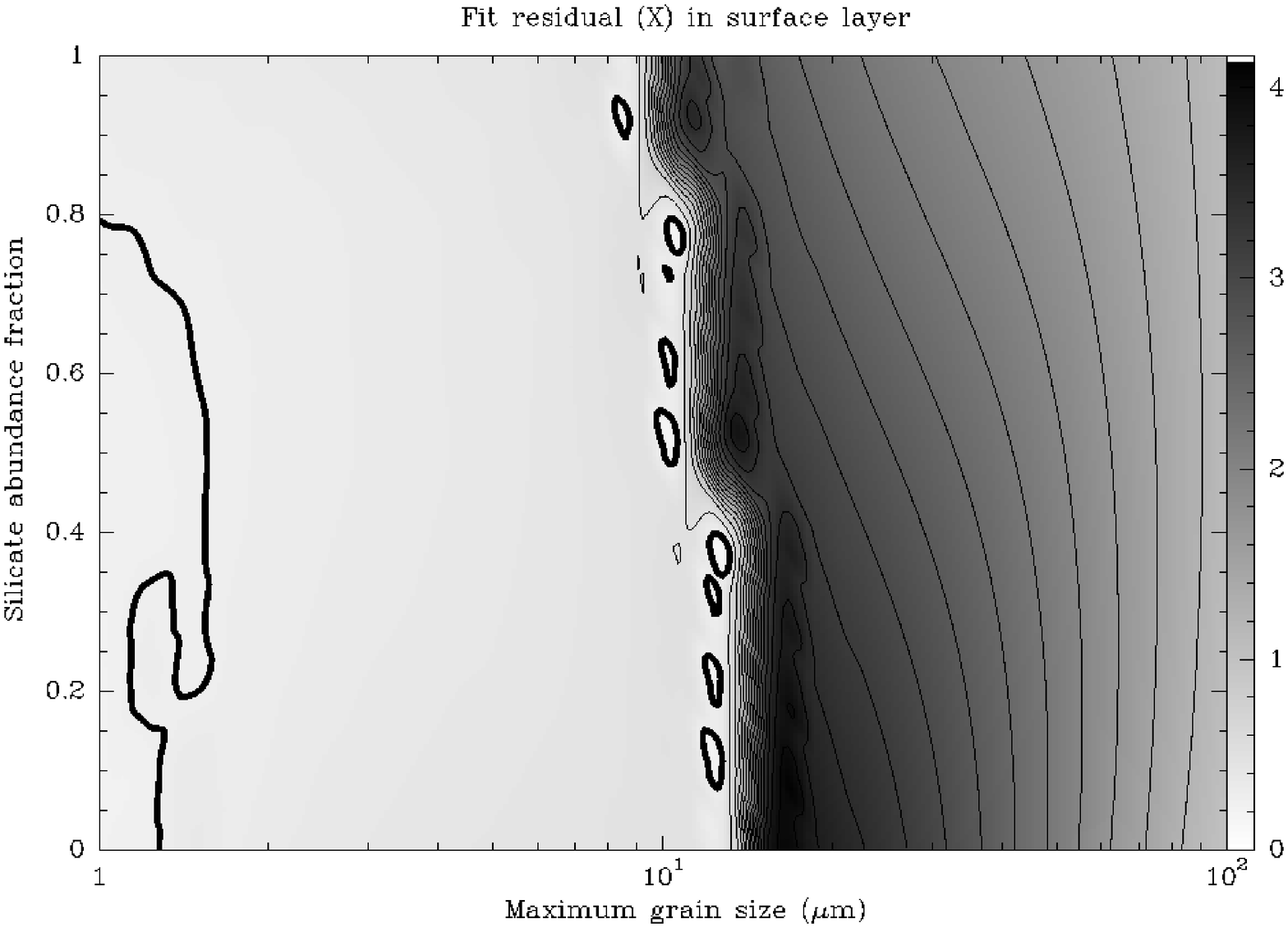}
\caption{Dispersion ($\chi$) as a function of the abundance of silicates and the maximum grain size for the interior (left) and surface (right) layers. Levels represent mean relative errors, starting from 0.25 in steps of 0.25. Acceptable fits can be considered for any values of the parameter space where the dispersion is below 0.25 (thick level).}
\label{fig:3}
\end{figure*}
\clearpage

\begin{table}
\caption{VV Ser model parameters. }
\begin{tabular}{lll}
\hline\noalign{\smallskip}
\textbf{Star parameters} & \textbf{Adopted value} & \textbf{References} \\
\noalign{\smallskip}\hline\noalign{\smallskip}
Spectral Type   & A0  & Mora et al. 2001 \\
Distance (pc)   & 260 & Straizys et al. 1996\\
Effective temperature (K) & 10000 & Pontoppidan et al. 2007 \\
Luminosity (L$_\odot$) & 49 & Pontoppidan et al. 2007 \\
\hline\noalign{\smallskip}
\textbf{Model parameters} & \textbf{Fitted value} & \textbf{Uncertainty}\tablenotemark{(a)} \\
\noalign{\smallskip}\hline\noalign{\smallskip}
Rim temperature (K) & 1500 & \tablenotemark{(b)} \\
Disk outer radius (AU) & 43 & $^{+25}_{-10}$\\
Disk mass (only dust, M$_\odot$) & 4~$10^{-5}$ & $\pm$ 3~$10^{-5}$ \\
Disk inclination (degrees) & 81 & $^{+9}_{-30}$ \\
Silicate abundace in interior layer & 0.36 & $\pm$ 0.4 \\
Silicate abundace in surface layer & 0.1 & $^{+0.7}_{-0.1}$ \\
Graphite abundace in interior layer & 0.64 & $\pm$ 0.4 \\
Graphite abundace in surface layer & 0.9 & $^{+0.7}_{-0.1}$ \\
Maximum grain size in interior layer ($\mu$m) & 10 000 & $\pm$ 1000 \\
Maximum grain size in surface layer ($\mu$m) & 1 & $^{+10}_{-0.5}$ \\
Grain size distribution slope (p) & 2.5 & \tablenotemark{(b)} \\
Density power law index & -1.97 & $\pm$ 0.2 \\
Disk rim radius (AU) & 0.53 & \tablenotemark{(c)} \\
\noalign{\smallskip}\hline
\end{tabular}
\label{tab:1b}       

\tablenotetext{(a)}{For $\chi$= 0.25, corresponding to a mean discrepancy of 25$\%$ between the observed and modeled fluxes.}
\tablenotetext{(b)}{Fixed parameter.}
\tablenotetext{(c)}{Obtained self-consistently from model.}

\end{table}
\clearpage

\section{Discussion}

\subsection{Near- and mid-infrared SED}
In Fig. 2a we show our fit to the SED towards VV~Ser. 
The fluxes at wavelengths shorter than 3~$\mu$m cannot be accounted by our model since it does not include the extinction of the stellar radiation by the surrounding dust. 
VV~Ser is a UX~Orionis star that presents flux variations in UV that are very likely to be because of spatially localized extinction events. To model these extinction events is beyond the scope of this Paper that is mainly centered on the mm-cm part of the SED.

We obtain a good fit in the near ($>$3~$\mu$m) and mid IR range with a slightly flared geometry, a rim radius $R_{rim}~=~0.53$~AU, and a disk scale height $H_{rim}/R_{rim}~=~0.17$. This geometry results in a partially (up to 6 AU) self-shadowed disk. A different (non-hydrostatic equilibrium) model was recently proposed by \cite{pon07} to account for the near-IR and mid-IR SED. These authors claimed a very complex geometry, a puffed-up inner rim formed by small grains (0.01--0.4 $\mu$m) at a radius $R_{rim}~=~0.8$~AU but allowing the larger grains in the midplane (0.3--3~$\mu$m) to penetrate to 0.25~AU, in an attempt to reconcile the interferometric measurements from \cite{eis07} with the optical light curve, and the disk shadow. Near-IR interferometry data are better fitted with R$_{rim}$$\sim$0.3--0.5~AU \citep{ise06,eis07}. However, in assuming Keplerian rotation and that the variations in the optical light curve were produced by extinction events in the inner rim, \cite{pon07} required R$_{rim} \sim$~0.8~AU to account for the width of the extinction events. Finally, providing that the grain dust mass is 0.8~10$^{-7}$~M$_\odot$, R$_{rim} \sim$~0.8~AU was required to account for the disk shadow.
All these facts prompted these authors to propose a different inner radius for the large grains in the mid-plane 
(responsible for the near-IR  interferometry results) from the small grains in the surface (responsible for
the extinction events and disk shadow).

Our model with $R_{rim}~=~0.53$~AU naturally accounts for the interferometric data. In addition, our millimeter
and centimeter data shows that the dust mass of the disk is two orders of magnitude larger than that
derived by \cite{pon07}. The shadow of the disk depends on the height at which the disk becomes optically
thick to UV photons, which is degenerate in the disk scale height and the total disk mass. An increment in the disk mass by two orders of magnitude would reduce the required value of the scale height and make it compatible with the value in hydrostatic equilibrium. 
Non-keplerian motions or a more complex geometry
of the inner rim could explain this discrepancy. As pointed above, our main goal is to explain the mm-cm SED and to derive the mass and properties of the large grains in the midplane. Since these parameters are quite independent on the detailed geometry of the inner rim, we prefer not to add more complexity to the model.

\subsection{Millimeter and centimeter SED}
The new mm and cm data allow us to improve the disk mass estimate and to constrain the dust properties. 
The low spectral index in the sub-mm--mm part can only be fitted with an opacity index $\beta\leq$0.3. This implies the existence of grains with sizes as large as 1~cm. These large grains have a very low emissivity in the whole range of frequencies and are unable to reproduce the large fluxes measured in the IR range. 
The only way to fit the whole spectrum is to assume a different grain population in the disk midplane 
from that in the disk surface and inner rim. We have allowed the grain properties to vary independently in the interior and surface layers to search for the best fit. The grain properties of these two populations can be determined quite independently since the emission in the near- and mid-IR range is mainly due to the surface layer and inner rim while at longer wavelengths the main contribution comes from the interior layer. 
The minimum dispersion is found for surface grains with a silicates abundance of 0.1 (which implies an abundance of 0.9 for graphite), a maximum grain size of 1~$\mu$m and a size distribution with $p~=~2.5$ for both grain components. The grain mixture is not well determined  since we obtain acceptable fits for a wide range of silicates abundance values (see Fig.~\ref{fig:3} and Table 2). However, the maximum grain size is well constrained to values between 1 and 10~$\mu$m. 
Our results are in agreement with previous model by \cite{ise06} that derived grain sizes of $>$1.2~$\mu$m in the inner rim, and a dust temperature of 1400~K.
The derived silicates abundance range and relatively large grain sizes are consistent with the flat 10~$\mu$m silicate feature observed in this source \citep{pon07}.



In the midplane we fit the observations with a silicate abundance of 0.36 (0.64 for graphite), a maximum grain size of 1~cm and 
a size distribution with $p~=~2.5$ for both grains (see Fig.~\ref{fig:1b}). Again, while the value of a$_{max}$ and $p$ is heavily constrained by the observations, the mixture of grains is poorly determined (see Fig.~\ref{fig:3} and Table 2). The fitted grain model provides a dust opacity $\kappa_{1.3mm}$= 0.2 cm$^2$ g$^{-1}$ and $\beta$=0.3. 
As shown by \cite{dra06}, the opacity index naturally becomes lower than unity for $p~=~3.5$ and a 
maximum grain size of 3~mm or larger (see Fig. 3). There is some degeneracy between the slope of the grain size distribution and the maximum grain size. A steeper grain size distribution, $p~=~3.0$ or 3.5, can be compensated by a larger value of the maximum grain size. A slope $p~=~2.5$ means that a larger number of grains are closer to the maximum grain size than for a more typical slope of 3.0-3.5. The values of $\kappa_{1.3mm}$ and $\beta$ that we obtain for p=2.5 and a$_{max}$=1~cm are very similar to those derived by \cite{dra06} for a maximum grain size of 10~cm and p = 3.5.
We have not explored the possibility of grains larger than 1~cm due to the extremely long computing 
time required to obtain the opacity of very big grains and consequently, our maximum grain size is actually a lower limit to the real one. 


The maximum optical depth (towards the star) computed from the model at 1 mm is 0.05. Since the emission is optically thin, the dust mass is proportional to the mm flux, and that can be estimated provided that the dust temperature and emissivity are known. We need a mass of 4 10$^{-5}$ M$_\odot$ to reproduce the mm fluxes. This mass is 400 times larger than previously 
predicted from the IR SED and shows that mass estimates based on far-IR fluxes can be wrong 
by several orders of magnitude. Measurements at mm wavelengths are required to have a reliable 
estimate of this parameter. The value of $\kappa_{1.3mm}$ we use, 0.2 cm$^2$ g$^{-1}$, is 5 times lower than the standard value $\kappa_{1.3mm}$~=~1.0~cm$^2$~g$^{-1}$ and consequently provides a disk mass 5 times larger. If the grain sizes were larger than a few cm, the value of $\kappa_{1.3mm}$ would be lower than 0.2 cm$^2$ g$^{-1}$ (see Fig. 3), and the disk mass would increase even more. However, the existence of grains larger than a few cm cannot be inferred from mm observations \citep{ale01}.

The radius of the disk is fitted to 43 AU, a value very similar to that found by \cite{pon07}. The size of the disk is very small compared to other disks detected around HAe stars whose radius ranges between 100~AU and 300~AU \citep{nat04, ise07}. Such a small size and mass makes the contribution from the inner rim to the observed flux to exceed that from the surface layer longwards of about 200~$\mu$m. This suggests that the disk could be truncated, possibly because of the existence of a secondary companion. However, no companion has been detected for this HAe star \citep{lei97, cor99}. Another possibility is that the outer part of the disk could be photo-evaporated due to the UV field of the star itself and other surrounding stars. \cite{tes98} derived the clustering degree in a wide sample of HAEBE stars. VV~Ser, surrounded by 24 IR stars in a radius of 0.2~pc, belongs to one of the most crowded clusters around any known HAe star.

\section{CO content of the disk}
We have not detected CO emission in the disk around VV~Ser down to a $\sigma$ of 14~mJy, or 0.3~K. Low abundances of CO are observed towards T Tauri stars and have been interpreted as the consequence of the CO freeze-out onto the grain surfaces in the dense and cold layers (T$_{dust}<$20~K) of the disk midplane. Assuming that the gas and dust are thermalized, depletion is not expected in the compact and warm disk around VV~Ser (T$_{dust}\geq$60~K all across the disk). One possibility is that the emission of the CO 1$\rightarrow$0 line is optically thick. This would be the case for a gas/dust ratio of 100 and a standard CO abundance. Our limit to the CO 1$\rightarrow$0 emission would then imply an upper limit to the disk size. Assuming a disk temperature of 200~K, the disk radius should be lower than 60~AU to be compatible with our non-detection. This radius is consistent with the radius derived from the model of dust emission. Another possibility is a low CO abundance because of photodissociation by the stellar UV radiation. In this case the CO emission could be optically thin and our limit would imply an upper limit for the CO abundance. Assuming that the CO emission is thermalized, a rotational temperature of 200~K and a typical linewidth of 15~km~s$^{-1}$, our limit would imply a CO abundance lower than 2.0~$10^{-6}$, i.e., 50 times lower than the standard value in the interstellar medium. Photodissociation of CO have been proposed to explain the lack of CO emission in other HAeBe stars \citep{man07}. Finally the lack of CO could be due to a low gas/dust mass ratio. We cannot distinguish between a low CO abundance and a low gas/dust mass ratio with observations of only one molecular species.

\section{Conclusions}

The main conclusion of this work is the presence of large grains in the midplane of the disk around VV~Ser. This result is supported by the flattish shape of the SED, and the low value of the spectral index between 1.3 and 2.6~mm ($\beta\leq$0.3). We have fitted the SED with a maximum grain size of 1~cm and $p~=~2.5$. As mentioned in previous sections the resulting grain size is in fact a lower limit to the actual size of the grains that could have grown to larger sizes. This is one of the still few disks towards which convincing evidences for the existence of cm-size particles ($``$pebbles$''$) have been found \citep{nat07}. In addition we find dust segregation in the vertical direction, with larger grains in the disk midplane and smaller ones in the disk surface. Evidence for a spatial separation between small and large grains has also been found in the transition disk around the HAe star IRS 48 \citep{gee07a}. Grain growth and dust settling are essential processes in any theory of planet formation. The detection and study of these processes in different disks are therefore required for the full comprehension of the formation of planetary systems. 



\acknowledgments
We thank the IRAM staff in Grenoble for their help and support during the observations and data reduction. We also thank Richard Dodson for his careful reading of the manuscript. T.A., A.F., R.B. and P.P. are grateful for support from the Spanish MEC under grant AYA 2003-07584. This work was partly supported by INAF PRIN 2006 grant "From disks to planetary systems".

\end{document}